\documentclass[nojss]{jss}\usepackage[]{graphicx}\usepackage[]{color}
\makeatletter
\def\maxwidth{ %
  \ifdim\Gin@nat@width>\linewidth
    \linewidth
  \else
    \Gin@nat@width
  \fi
}
\makeatother

\usepackage{amsmath,bm}
\usepackage{subcaption,fancyvrb}

\usepackage{tikz}
\usetikzlibrary{decorations.pathmorphing,backgrounds,fit,calc,through,arrows,shapes,trees,positioning,matrix,chains}
\tikzset {crossline/.style={opacity=.15,line width=4.5mm,line cap=round,color=#1},
	method/.style={opacity=.15,fill,rounded corners,color=blue},
	manifest/.style={rectangle,draw,inner sep=0.15em,minimum size=2.5em},
	latent/.style={circle,draw,inner sep=0.15em,minimum size=2.7em},
	error/.style={circle,draw,inner sep=0.15em,minimum size=2em},
        errorsling/.style={<->,shorten >=1pt,>=stealth',shorten <=1pt,>=stealth',auto,distance=8mm,semithick},
	every loop/.style={min distance=1em,in=70,out=110,looseness=3.5},
	highlight/.style={rectangle,rounded corners,fill=red!15,draw,fill opacity=0.5,thick,inner sep=0pt}
}

\author{Edgar C.\ Merkle\\University of Missouri \And 
        Yves Rosseel\\Ghent University}
\title{\pkg{blavaan}: Bayesian Structural Equation Models via Parameter Expansion}

\Plainauthor{Edgar C.\ Merkle, Yves Rosseel} %% comma-separated
\Plaintitle{Bayesian Structural Equation Models} %% without formatting
\Shorttitle{Bayesian SEM} %% a short title (if necessary)

\Abstract{
This article describes \pkg{blavaan}, an \proglang{R} package for estimating Bayesian structural equation models (SEMs) via \proglang{JAGS} and for summarizing the results. It also describes a novel parameter expansion approach for estimating specific types of models with residual covariances, which facilitates estimation of these models in \proglang{JAGS}. The methodology and software are intended to provide users with a general means of estimating Bayesian SEMs, both classical and novel, in a straightforward fashion.  Users can estimate Bayesian versions of classical SEMs with \pkg{lavaan} syntax, they can obtain state-of-the-art Bayesian fit measures associated with the models, and they can export \proglang{JAGS} code to modify the SEMs as desired.  These features and more are illustrated by example, and the parameter expansion approach is explained in detail.
}
\Plainkeywords{Bayesian SEM, structural equation models, JAGS, MCMC, lavaan} %% without formatting
\Address{
  Edgar C.\ Merkle\\
  Department of Psychological Sciences\\
  University of Missouri\\
  28A McAlester Hall\\
  Columbia, MO, USA 65211\\
  E-mail: \email{merklee@missouri.edu}\\
  URL: \url{http://faculty.missouri.edu/~merklee/}
}
\IfFileExists{upquote.sty}{\usepackage{upquote}}{}
\begin{document}

%% include your article here, just as usual
%% Note that you should use the \pkg{}, \proglang{} and \code{} commands.

\section{Introduction}

The intent of \pkg{blavaan} is to implement Bayesian structural equation models (SEMs) that harness open source MCMC samplers \citep[in \proglang{JAGS};][]{plu03} while simplifying model specification, summary, and extension. 
Bayesian SEM has received increasing attention in recent years, with MCMC samplers being developed for specific priors \citep[e.g.,][]{lee07,schhoi99}, 
specific models being estimated in \proglang{JAGS} and \proglang{BUGS} \citep{bugs12,luntho00}, 
factor analysis samplers being included in R packages \pkg{bfa} \citep{bfa} and \pkg{MCMCpack} \citep{mcmcpack},
and multiple samplers being implemented in \proglang{Mplus} \citep{mutasp12}. These methods have notable advantages over analogous frequentist methods, including the facts that estimation of complex models is typically easier \citep[e.g.,][]{marwen13} and that estimates of parameter uncertainty do not rely on asymptotic arguments. Further, in addition to allowing for the inclusion of existing knowledge into model estimates, prior distributions can be used to inform underidentified models, to average across multiple models in a single run \citep[e.g.,][]{lucho16}, and to avoid Heywood cases \citep{marmcd75}.

The traditional SEM framework \citep[e.g.,][]{bol89} assumes an $n \times p$ data matrix $\bm{Y} = (\bm{y}_1\ \bm{y}_2\ \ldots\ \bm{y}_n)^\top$ with $n$ independent cases and
$p$ continuous manifest variables.  Using the LISREL ``all y'' notation \citep[e.g.,][]{jorsor97}, a structural equation model
with $m$ latent variables may be represented by the equations
\begin{align}
  \label{eq:sem1}
    \bm{y} &= \bm{\nu} + \bm{\Lambda} \bm{\eta} +
    \bm{\epsilon} \\
  \label{eq:sem2}
    \bm{\eta} &= \bm{\alpha} + \bm{B}\bm{\eta} + \bm{\zeta},
\end{align}
where $\bm{\eta}$ is an $m \times 1$ vector containing the latent 
variables; $\bm{\epsilon}$ is a $p \times 1$ vector of residuals; and $\bm{\zeta}$ is an $m \times 1$ vector of
residuals associated with the latent variables.  Each entry in these three vectors
is independent of the others.  Additionally, $\bm{\nu}$ and $\bm{\alpha}$ contain intercept parameters for the manifest and latent variables, respectively; $\bm{\Lambda}$ is a
matrix of factor loadings; and
$\bm{B}$ contains parameters
that reflect directed paths between latent variables (with the assumption that $(\bm{I} - \bm{B})$ is invertible).

In conventional SEMs, we assume multivariate normality of the $\bm{\epsilon}$ and $\bm{\zeta}$ vectors.  In particular,
\begin{align}
  \label{eq:mres}
  \bm{\epsilon} &\sim N_p(\bm{0}, \bm{\Theta}) \\
  \label{eq:lres}
  \bm{\zeta} &\sim N_m(\bm{0}, \bm{\Psi}),
\end{align}
with the latter equation implying multivariate normality of the latent
variables.  Taken together, the above assumptions imply that the marginal distribution of $\bm{y}$ (integrating out the latent variables) is multivariate normal with parameters
\begin{align}
\bm{\mu} &= \bm{\nu} + \bm{\Lambda \alpha} \\
\bm{\Sigma} &= \bm{\Lambda}(\bm{I} - \bm{B})^{-1} \bm{\Psi} (\bm{I} - \bm{B}^\top)^{-1} \bm{\Lambda}^\top + \bm{\Theta}.
\end{align}

If one is restricted to conjugate priors, then the \cite{sonlee12} or \cite{mutasp12} procedures are often available for simple MCMC estimation of the above model. However, if one wishes to use general priors or to estimate novel models, then there is the choice of implementing a custom MCMC scheme or using a general MCMC program like \proglang{BUGS}, \proglang{JAGS}, or \proglang{Stan} \citep{stan14}. These options are often time-consuming and difficult to extend to further models. This is where \pkg{blavaan} is intended to be helpful: it allows for simple
specification of Bayesian SEMs while also allowing the user to extend the original model.  
In addition to easing Bayesian SEM specification, the package includes a novel approach to \proglang{JAGS} model estimation that allows us to handle models with correlated residuals. This approach, further described below, builds on the work
of many previous researchers: the approach of \cite{lee07} for estimating models in \proglang{WinBUGS}; the approach of \cite{paldun07} for handling correlated residuals; and the approaches of \cite{barmcc00} and \cite{mutasp12} for specifying prior distributions of covariance matrices.

Package \pkg{blavaan} is potentially useful for the analyst that wishes to estimate Bayesian SEMs and for the methodologist that wishes to develop novel extensions of Bayesian SEMs.  The package is a series of bridges between several
existing \proglang{R} packages, with the bridges being used to make each step of the
modeling easy and flexible.  Package \pkg{lavaan} \citep{ros12} serves as the starting point and ending
point in the sequence: the user specifies models via \pkg{lavaan} syntax, and objects of class \pkg{blavaan} make use of many \pkg{lavaan} functions.  Consequently, writings on \pkg{lavaan} \citep[e.g.,][]{bea14,ros12}
give the user a good idea about how to use \pkg{blavaan}.

Following
model specification, \pkg{blavaan} examines the details of
the specified model and converts the specification to \proglang{JAGS}
syntax.  Importantly, this conversion often makes use of the parameter
expansion ideas presented below, resulting in MCMC
chains that quickly converge for many models.  We employ package
\pkg{runjags} \citep{den15} for sampling parameters and summarizing the MCMC chains.
Once \pkg{runjags} is finished, \pkg{blavaan} organizes summaries and computes a variety of Bayesian fit measures.

\section{Bayesian SEM}

% describe lee approach, conjugate priors
As noted above, the approach of Lee \citep[\citeyear{lee07}; see also][]{sonlee12} involves the use of conjugate priors on the
parameters from Equations~\eqref{eq:sem1} to~\eqref{eq:lres}. Specific priors include inverse gamma distributions on variance parameters, inverse Wishart distributions on unrestricted covariance matrices (typically covariances of exogenous latent variables), and normal distributions on other parameters. 

Importantly, Lee assumes that the manifest variable covariance matrix $\bm{\Theta}$ and the endogenous latent variable covariance matrix $\bm{\Psi}$ are diagonal. This
assumption of diagonal covariance matrices is restrictive in some
applications.  For example, researchers often wish to fit models with
correlated residuals, and it has been argued that such models are both
necessary and under-utilized \citep{colcie07}.  Correlated residuals
pose a difficult problem for MCMC methods because they often result in
covariance matrices with some (but not all) off-diagonal entries
equal to zero. 
In this situation, we cannot assign an inverse Wishart prior to the
full covariance matrix because this does not allow us to fix some
off-diagonal entries to zero.  Further, if we assign a prior to each
free entry of the covariance matrix and sample the parameters
separately, we can often obtain covariance matrices that are not
positive definite. This can lead to numerical problems, resulting in an inability to carry out model estimation.
Finally, it is possible to specify an equivalent model that possesses the required, diagonal matrices. However, the setting of prior distributions can be unclear here: if the analyst specifies prior distributions for her model of interest, then it may be cumbersome to translate these into prior distributions for the equivalent model.

To address the issue of non-diagonal covariance matrices,
\cite{mutasp12} implemented a random walk method that is based on work
by \cite{chigre98}.  This method samples free
parameters of the covariance matrix via Metropolis-Hastings steps.
While the implementation is fast and efficient, it does not allow
for some types of equality constraints because parameters are
updated in blocks (either all parameters in a block must be constrained to equality, or no parameters in a block can be constrained).  Further, the method is unreliable for models
involving many latent variables.  Consequently, \cite{mutasp12}
implemented three other MCMC methods that are suited to different
types of models.

In our initial work on package \pkg{blavaan}, we developed methodology
for fitting models with free residual covariance parameters. We sought
methodology that (i) would work in \proglang{JAGS}, (ii) was
reasonably fast and efficient (by \proglang{JAGS} standards), and (iii) allowed for satisfactory specification of
prior distributions.  In the
next section, we describe the resulting methodology.  The methodology can often be used to reliably estimate posterior means to the first decimal place on the order
of minutes (often, less than two minutes on desktop computers, but it also depends on model complexity).  This will never beat compiled code, but it is typically better than
alternative \proglang{JAGS} parameterizations that rely on
multivariate distributions.

\subsection{Parameter expansion}

General parameter expansion approaches to Bayesian inference are
described by \cite{gel04,gel06}, with applications to
factor analysis models detailed by \cite{ghodun09}.
Our approach here is related to that of \cite{paldun07}, who employ phantom latent variables (they use the term pseudo-latent variable) to simplify the estimation of models with non-diagonal $\bm{\Theta}$ matrices.  This results in a {\em working model} that is estimated, with the working model parameters being transformed to the {\em inferential model} parameters from Equations~\eqref{eq:sem1} and~\eqref{eq:sem2}. The use of phantom latent variables in SEM has long been known \citep[e.g.,][]{rin84}, though the Bayesian approach involves new issues associated with prior distribution specification. This is further described below.

\subsubsection{Overview}

Assuming $v$ nonzero entries in the lower triangle of $\bm{\Theta}$ (i.e., $v$ residual covariances), \cite{paldun07} take the inferential model residual vector $\bm{\epsilon}$ and reparameterize it as
\begin{align*}
  \bm{\epsilon} &= \bm{\Lambda}_D \bm{D} + \bm{\epsilon}^* \\
  \bm{D} &\sim N_v(\bm{0}, \bm{\Psi}_D) \\
  \bm{\epsilon}^* &\sim N_p(\bm{0}, \bm{\Theta}^*),
\end{align*}
where $\bm{\Lambda}_D$ is a $p \times v$ matrix with many zero
entries, $\bm{D}$ is a $v \times 1$ vector of phantom latent variables, and $\bm{\epsilon}^*$ is a $p \times 1$ residual vector.  This approach is useful because, by carefully choosing the nonzero entries of $\bm{\Lambda}_D$, both $\bm{\Psi}_D$ and $\bm{\Theta}^*$ are diagonal.  This allows us to employ an approach related to \cite{lee07}, avoiding high-dimensional normal distributions in favor of univariate normals.  

Under this working model parameterization, the inferential model covariance matrix $\bm{\Theta}$ can be re-obtained via
\begin{equation*}
  \bm{\Theta} = \bm{\Lambda}_D \bm{\Psi}_D \bm{\Lambda}_D^\top + \bm{\Theta}^*.
\end{equation*}

We can also handle covariances between latent variables in the same way, as necessary. Assuming $m$ latent variables with $w$ covariances, we have
\begin{align*}
  \bm{\zeta} &= \bm{B}_E \bm{E} + \bm{\zeta}^* \\
  \bm{E} &\sim N_w(\bm{0}, \bm{\Psi}_E) \\
  \bm{\zeta}^* &\sim N_m(\bm{0}, \bm{\Psi}^*),
\end{align*}
where the original covariance matrix $\bm{\Psi}$ is re-obtained via:
\begin{equation*}
  \bm{\Psi} = \bm{B}_E \bm{\Psi}_E \bm{B}_E^\top + \bm{\Psi}^*.
\end{equation*}
This approach to handling latent
variable covariances can lead to slow convergence in models
with many (say, $>5$) correlated latent variables.  In these cases, we have
found it better (in \proglang{JAGS}) to sample the latent variables from multivariate normal
distributions.  Package \pkg{blavaan} attempts to use multivariate
normal distributions in situations where it can, reverting to the
above reparameterization only when necessary.

To choose nonzero entries of $\bm{\Lambda}_D$ (with the same method
applying to nonzero entries of $\bm{B}_E$), we
define two $v \times 1$ vectors $\bm{r}$ and $\bm{c}$.  These vectors
contain the respective row numbers and column numbers of the nonzero,
lower-triangular entries of $\bm{\Theta}$.  For $j=1,\ldots,v$, the
nonzero entries of $\bm{\Lambda}_D$ then occur in column $j$, rows
$r_j$ and $c_j$.  \cite{paldun07} set these nonzero entries equal to 1,
which can be problematic if the covariance parameters' posteriors are negative or overlap with zero. This is because the only remaining free parameters are variances, which can only be positive. Instead of 1s, we free the parameters in 
$\bm{\Lambda}_D$ so that they are functions of inferential model parameters.  This is further described in the next section.  First, however, we give an example of the approach.

\subsubsection{Example}
\begin{figure}
\begin{Schunk}

{\centering \includegraphics[width=4.5in,height=9in]{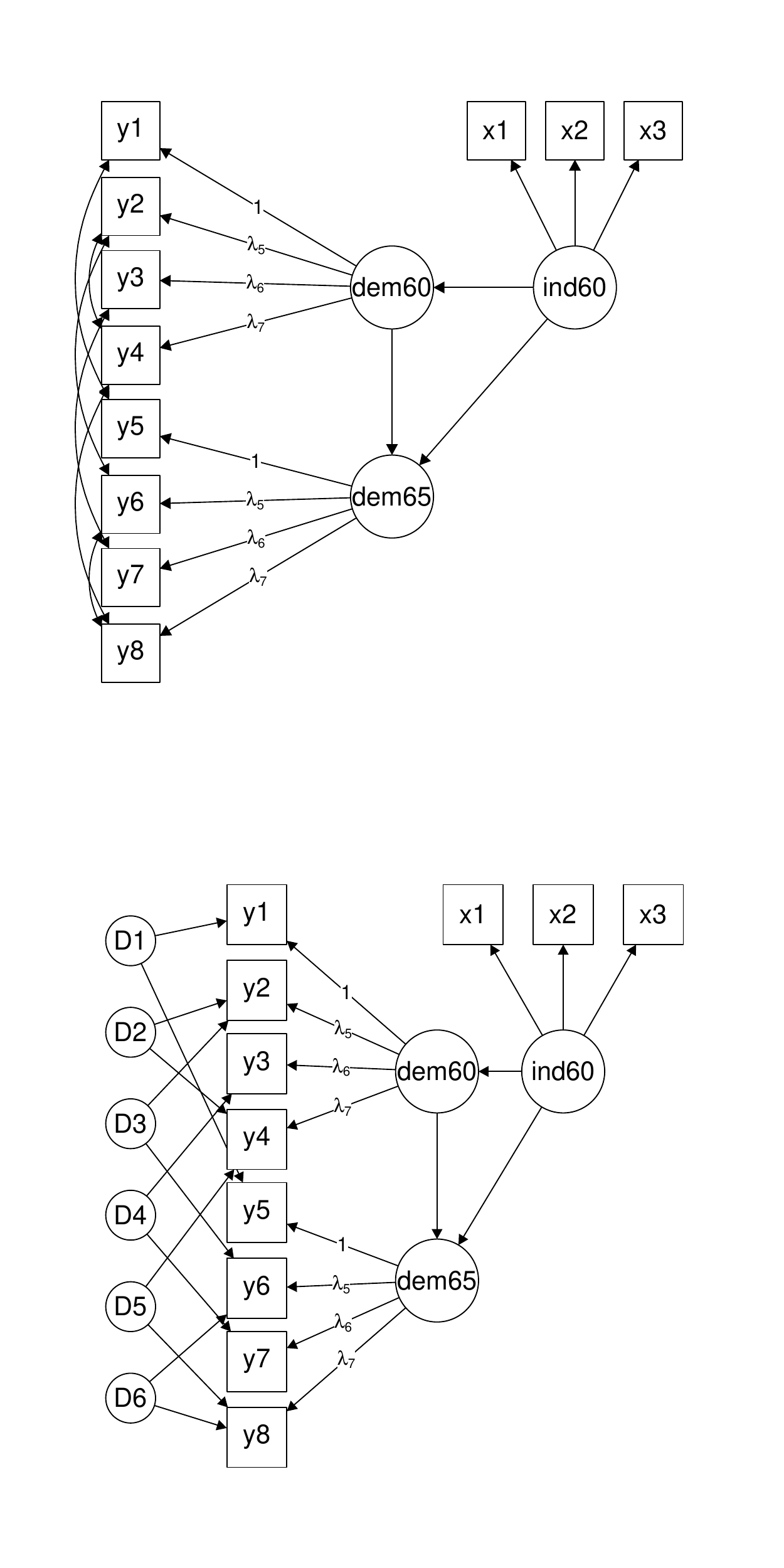} 

}

\end{Schunk}
\caption{Political democracy path diagrams, original (top) and
  parameter-expanded version (bottom). Variance parameters are omitted.}
\label{fig:pathpd}
\end{figure}

Consider the political democracy example of \cite{bol89}, which
includes eleven observed variables that are hypothesized to arise from
three latent variables.  The inferential structural equation model
affiliated with these data appears as a path diagram (with variance
parameters omitted) at the top of Figure~\ref{fig:pathpd}.  
This model can be written in matrix form as

{\scriptsize
\begin{align*}
  \begin{pmatrix} y_1 \\ y_2 \\ y_3 \\ y_4 \\ y_5 \\
          y_6 \\ y_7 \\ y_8 \\ y_9 \\ y_{10} \\ y_{11} \end{pmatrix}
      &= \begin{pmatrix} \nu_1 \\ \nu_2 \\ \nu_3 \\ \nu_4 \\ \nu_5 \\ \nu_6
          \\ \nu_7 \\ \nu_8 \\ \nu_9 \\ \nu_{10} \\ \nu_{11} \end{pmatrix} + 
         \begin{pmatrix} 1 & 0 &
              0 \\ \lambda_2 & 0 & 0 \\ \lambda_3 & 0 & 0 \\ 0 & 1 & 0 \\
              0 & \lambda_5 & 0 \\ 0 & \lambda_6 & 0 \\ 0 & \lambda_7
              & 0 \\ 0 & 0 & 1 \\ 0 & 0 & \lambda_5 \\ 0 & 0 &
              \lambda_6 \\ 0 & 0 &
              \lambda_7 \end{pmatrix} \begin{pmatrix} \text{ind60} \\
              \text{dem60} \\ \text{dem65} \end{pmatrix} +
              \begin{pmatrix} e_1 \\ e_2 \\ e_3 \\ e_4 \\ e_5 \\
          e_6 \\ e_7 \\ e_8 \\ e_9 \\ e_{10} \\ e_{11} \end{pmatrix} \\
          & \\
  \begin{pmatrix} \text{ind60} \\ \text{dem60} \\
      \text{dem65} \end{pmatrix} &= \begin{pmatrix} 0 \\ 0 \\ 0 \end{pmatrix} + 
      \begin{pmatrix} 0 & 0 & 0 \\ b_1 & 0 & 0 \\ b_2 & b_3 & 0 \end{pmatrix} 
      \begin{pmatrix} \text{ind60} \\ \text{dem60} \\ \text{dem65} \end{pmatrix} + \begin{pmatrix} \zeta_1 \\ \zeta_2 \\ \zeta_3 \end{pmatrix}.
\end{align*}}

Additionally, the latent residual covariance matrix $\bm{\Psi}$ is
diagonal and the observed residual covariance matrix is

\setcounter{MaxMatrixCols}{13}
{\scriptsize
\begin{equation*}
    \bm{\Theta} = \begin{pmatrix}
\theta_{11} &  &  &  &  &  &  &  &  &  &  \\
 & \theta_{22} &  &  &  &  &  &  &  &  &  \\
 &  & \theta_{33} &  &  &  &  &  &  &  &  \\
 &  &  & \theta_{44} &  &  &  & \theta_{48} &  &  &  \\
 &  &  &  & \theta_{55} &  & \theta_{57} &  & \theta_{59} &  &  \\
 &  &  &  &  & \theta_{66} &  &  &  & \theta_{6,10} &  \\
 &  &  &  & \theta_{57} &  & \theta_{77} &  &  &  & \theta_{7,11} \\
 &  &  & \theta_{48} &  &  &  & \theta_{88} &  &  &  \\
 &  &  &  & \theta_{59} &  &  &  & \theta_{99} &  & \theta_{9,11} \\
 &  &  &  &  & \theta_{6,10} &  &  &  & \theta_{10,10} &  \\
 &  &  &  &  &  & \theta_{7,11} &  & \theta_{9,11} &  & \theta_{11,11} \\

\end{pmatrix},
\end{equation*}}
\vspace{.2in} \  \\
where the blank entries all equal zero.
The six unique, off-diagonal entries of $\bm{\Theta}$ make
the model difficult to estimate via Bayesian methods.

To ease model estimation, we follow
\cite{paldun07} in reparameterizing the model to the working model
displayed at the bottom of Figure~\ref{fig:pathpd}.  The working model
is now

{\scriptsize
\begin{align*}
  \begin{pmatrix} y_1 \\ y_2 \\ y_3 \\ y_4 \\ y_5 \\
          y_6 \\ y_7 \\ y_8 \\ y_9 \\ y_{10} \\ y_{11} \end{pmatrix}
      &= \begin{pmatrix} \nu_1 \\ \nu_2 \\ \nu_3 \\ \nu_4 \\ \nu_5 \\ \nu_6
          \\ \nu_7 \\ \nu_8 \\ \nu_9 \\ \nu_{10} \\ \nu_{11} \end{pmatrix} + 
         \begin{pmatrix} 1 & 0 &
              0 \\ \lambda_2 & 0 & 0 \\ \lambda_3 & 0 & 0 \\ 0 & 1 & 0 \\
              0 & \lambda_5 & 0 \\ 0 & \lambda_6 & 0 \\ 0 & \lambda_7
              & 0 \\ 0 & 0 & 1 \\ 0 & 0 & \lambda_5 \\ 0 & 0 &
              \lambda_6 \\ 0 & 0 &
              \lambda_7 \end{pmatrix} \begin{pmatrix} \text{ind60} \\
              \text{dem60} \\ \text{dem65} \end{pmatrix} +
          \begin{pmatrix} 0 & 0 & 0 & 0 & 0 & 0 \\
                          0 & 0 & 0 & 0 & 0 & 0 \\
                          0 & 0 & 0 & 0 & 0 & 0 \\
                          \lambda_{D1} & 0 & 0 & 0 & 0 & 0 \\ 
                          0 & \lambda_{D2} & \lambda_{D3} & 0 & 0 & 0 \\ 
                          0 & 0 & 0 & \lambda_{D4} & 0 & 0 \\ 
                          0 & \lambda_{D5} & 0 & 0 & \lambda_{D6} & 0 \\
                          \lambda_{D7} & 0 & 0 & 0 & 0 & 0 \\ 
                          0 & 0 & \lambda_{D8} & 0 & 0 & \lambda_{D9} \\ 
                          0 & 0 & 0 & \lambda_{D10} & 0 & 0 \\
                          0 & 0 & 0 & 0 & \lambda_{D11} & \lambda_{D12} \end{pmatrix} 
          \begin{pmatrix} D1 \\ D2 \\ D3 \\ D4 \\ D5 \\
              D6 \end{pmatrix} + 
            \begin{pmatrix} e^*_1 \\ e^*_2 \\ e^*_3 \\ e^*_4 \\ e^*_5 \\
          e^*_6 \\ e^*_7 \\ e^*_8 \\ e^*_9 \\ e^*_{10} \\ e^*_{11} \end{pmatrix} \\
  \nonumber        & \\
  \begin{pmatrix} \text{ind60} \\ \text{dem60} \\
      \text{dem65} \end{pmatrix} &= \begin{pmatrix} 0 \\ 0 \\ 0 \end{pmatrix} + 
      \begin{pmatrix} 0 & 0 & 0 \\ b_1 & 0 & 0 \\ b_2 & b_3 & 0 \end{pmatrix} 
      \begin{pmatrix} \text{ind60} \\ \text{dem60} \\ \text{dem65} \end{pmatrix} + 
      \begin{pmatrix} \zeta_1 \\ \zeta_2 \\ \zeta_3 \end{pmatrix}.
\end{align*}}

The covariance matrix associated with $\bm{\epsilon}^*$,
$\bm{\Theta}^*$, is now diagonal, which makes the model easier to
estimate via MCMC. The latent variable residual, $\bm{\zeta}$, is maintained as before because its covariance matrix was already diagonal. In general, we only reparameterize residual covariance matrices that are neither diagonal nor unconstrained. 

As mentioned earlier, the difference
between our approach and that of \cite{paldun07} is that we estimate
the loadings with $D$ subscripts, whereas \cite{paldun07} fix these loadings to 1.
This allows us to obtain posteriors on residual covariances that overlap with zero or that become negative, as necessary.
Estimation of these loadings comes with a cost, however, in that the prior distributions of the inferential model do not immediately translate into prior distributions of the working model. 
This problem and a solution are further described in the next section.

\subsection{Priors on covariances}

Due to the reparameterization of covariances, prior
distributions on the working model parameters require care in their specification.  Every inferential model covariance
parameter is potentially the product of three parameters: two representing paths
from a phantom latent variable to the variable of interest, and one
precision (variance) of the phantom latent variable.  
These three parameters also impact the inferential model variance parameters.
We carefully restrict these parameters to arrive at prior
distributions that are both meaningful and flexible.

We place univariate prior distributions on model variance and
correlation parameters, with the distributions being based on the inverse Wishart or on other prior distributions described in the literature.  This is highly related to the approach of \cite{barmcc00}, who separately specify prior distributions on the variance and correlation parameters comprising a covariance matrix.  Our approach is also related to \cite{mutasp12}, who focus on marginal distributions associated with the inverse Wishart.  A notable difference is that we always use proper prior distributions,
whereas \cite{mutasp12} often use improper prior distributions.

\begin{figure}
  \centering
  \begin{subfigure}[t]{0.4\textwidth}
    \centering
    \captionsetup{justification=centering,font=scriptsize}
    \begin{tikzpicture}
  \matrix[row sep=2.5mm, ampersand replacement=\&]
  {
    \node(X1)[manifest]{$X_1$} edge[errorsling, bend right=80]node[above left=.1cm and .45cm,text width=0.1cm]{$\scriptsize \theta_{11}$} (X1.170); \& \&[3em]
    \node(X2)[manifest]{$X_2$} edge[errorsling, bend right=80]node[above right=.1cm and .05cm,text width=0.1cm]{$\scriptsize \theta_{22}$} (X2.80); \\
  };

\draw[stealth'-stealth'] (X1.north east) to [out=70, in=110] (X2.north west)node[above left=.5cm and .6cm,text width=.1cm]{$\scriptsize \theta_{12}$};
\end{tikzpicture}
    \caption*{Inferential model}
  \end{subfigure}
  ~
  \begin{subfigure}[t]{0.4\textwidth}
    \centering
    \captionsetup{justification=centering,font=scriptsize}
    \resizebox{0.7\textwidth}{!}{
      \begin{tikzpicture}
  \matrix[row sep=1.5mm, ampersand replacement=\&]
  {
    \node(X1)[manifest]{$X_1$} edge[errorsling, bend right=80]node[above left=.15cm and .5cm,text width=0.1cm]{$\scriptsize \theta^{*}_{11}$} (X1.170); \& \&[4em]
    \node(X2)[manifest]{$X_2$} edge[errorsling, bend right=80]node[above right=.15cm and .05cm,text width=0.1cm]{$\scriptsize \theta^{*}_{22}$} (X2.80); \\
  };

  \node(P1)[latent] at ($(X1) + (3em,6em)$){$D$}
    edge[errorsling, bend right=80]node[above left=.05mm,text width=0.1cm]{$\scriptsize \psi_D$} (P1.130);

  \draw[-stealth'] (P1)--(X1.north) node[rectangle,pos=.4,fill=white,inner sep=0.3em]{$\scriptsize \lambda_1$};
  \draw[-stealth'] (P1)--(X2.north) node[rectangle,pos=.4,fill=white,inner sep=0.3em]{$\scriptsize \lambda_2$};

\end{tikzpicture}
    }
    \caption*{Working model}
  \end{subfigure}
  \caption{Example of phantom latent variable approach to covariances.}
  \label{fig:phanex}
\end{figure}
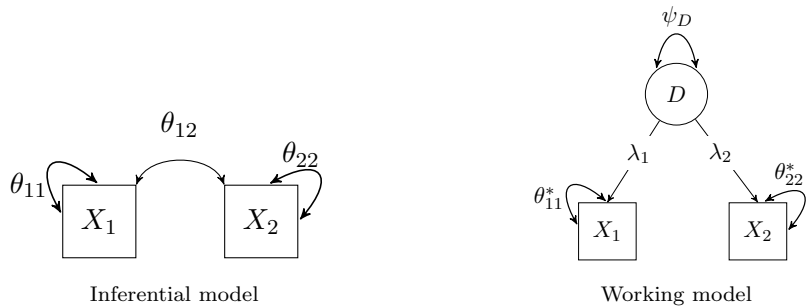

To illustrate the approach, we refer to Figure~\ref{fig:phanex}.  This
figure uses path diagrams to display two observed variables with
correlated residuals (left panel) along with their parameter-expanded
representation in the working model (right panel).  
Our goal is to specify a sensible prior distribution for the inferential
model's variance/covariance parameters in the left panel, using the parameters in the right panel.  
The covariance of
interest, $\theta_{12}$, has been converted to a 
normal phantom latent variable with variance $\psi_D$ and two
directed paths, $\lambda_1$ and 
$\lambda_2$.  To implement an approach related to that of \cite{barmcc00} in \proglang{JAGS}, we set 
\begin{align*}
    \psi_D &= 1 \\
    \lambda_1 &= \sqrt{|\rho_{12}|\theta_{11}} \\
    \lambda_2 &= \text{sign}(\rho_{12})\sqrt{|\rho_{12}|\theta_{22}} \\
    \theta^{*}_{11} &= \theta_{11} - |\rho_{12}|\theta_{11} \\
    \theta^{*}_{22} &= \theta_{22} - |\rho_{12}|\theta_{22},
\end{align*}
where $\rho_{12}$ is the correlation associated with the covariance $\theta_{12}$, and $\text{sign}(a)$ equals 1 if $a>0$ and $-1$ otherwise.  

Using this combination of parameters, we need only set prior distributions on the inferential model parameters $\theta_{11}$, $\theta_{22}$, and $\rho_{12}$, with $\theta_{12}$ being obtained from these parameters in the usual way.  If we want our priors to be analogous to an inverse Wishart prior with $d$ degrees of freedom and diagonal scale matrix $\bm{S}$, we can set univariate priors of
\begin{align*}
  \theta_{11} &\sim IG((d - p + 1)/2, s_{11}/2) \\
  \theta_{22} &\sim IG((d - p + 1)/2, s_{22}/2) \\
  \rho_{12} &\sim \text{Beta}_{(-1,1)}((d-p+1)/2, (d-p+1)/2),
\end{align*}
where $\text{Beta}_{(-1,1)}$ is a beta distribution with support on $(-1,1)$ instead of the usual $(0,1)$ and $p$ is the dimension of $\bm{S}$ (typically, the number of observed variables).
These priors are related to those used by \cite{mutasp12}, based on
results from \cite{barmcc00}.  They are also the default priors for
variance/correlation parameters in \pkg{blavaan}, with $d=(p+1)$ and
$\bm{S}=\bm{I}$.  We refer to this parameterization option as
\code{"srs"}, reflecting the fact that we are dissecting the
covariance parameters into standard deviation and
correlation parameters.

\cite{barmcc00} avoid the inverse gamma priors on the variance
parameters, instead opting for log-normal distributions on the standard deviations that they
judged to be more numerically stable.  Package \pkg{blavaan} allows for custom priors, so that the user can employ the log-normal or others on precisions, variances, or standard deviations (this is illustrated later).
This approach is generally advantageous because it allows for flexible specification of prior distributions on covariance matrices, including those with fixed zeros or those where we have different amounts of prior information on certain variance/correlation parameters within the matrix.

While we view the \code{"srs"} approach as optimal for dealing with
covariance parameters here, there exist similar alternative approaches in
\proglang{JAGS}.  For example, we can treat the phantom latent
variables similarly to the other latent variables in the model,
assigning prior distributions 
in the same manner.  This alternative approach, which we call \code{"fa"} (because
the priors resemble traditional factor analysis priors), involves the
following default prior distributions on the working model from Figure~\ref{fig:phanex}:
\begin{align*}
    \psi_D &\sim IG(1,.5) \\
    \lambda_1 &\sim N(0, 10^4) \\
    \lambda_2 &\sim N(0, 10^4) \\
    \theta_{11}^* &\sim IG(1,.5) \\
    \theta_{22}^* &\sim IG(1,.5),
\end{align*}
with the inferential model parameters being obtained in the usual way:
\begin{align*}
    \theta_{11} &= \theta^*_{11} + \lambda_1^2\psi_D \\
    \theta_{22} &= \theta^*_{22} + \lambda_2^2\psi_D \\
    \theta_{12} &= \lambda_1\lambda_2\psi_D.
\end{align*}
The main disadvantage of the \code{"fa"} 
approach is that the implied prior distributions on the inferential
model parameters are not of a common form.  Thus, it is difficult to
introduce informative prior distributions for the inferential model
variance/covariance parameters.  
For example, the prior on the inferential covariance ($\theta_{12}$) is the product of two normal prior distributions and an inverse gamma, which can be surprisingly informative. To avoid confusion here, we do not allow the user to directly modify the priors on the working parameters $\lambda_1$, $\lambda_2$, and $\psi_D$ under the \code{"fa"} approach. The priors chosen above are approximately noninformative for most applications, and the user can further modify the exported \proglang{JAGS} code if he/she desires.
Along with the prior distribution issue, the working model
parameters are not identified by the likelihood because each inferential covariance parameter is the product of the three working parameters. This can complicate Laplace approximation of the marginal likelihood (which, as described below, is related to the Bayes factor), because the approximation
works best when the posterior distributions are unimodal.

In \pkg{blavaan}, the user is free to specify \code{"srs"} or \code{"fa"}
priors for all covariance parameters in the model. In the example section, we present a simple comparison of these two options' relative speeds and efficiencies.
Beyond these options, the package attempts to identify when it can sample latent variables directly from a multivariate normal distribution, which can improve sampling efficiency.
This most commonly happens when we have many exogenous latent
variables that all covary with one another.  In this case, we place an
inverse Wishart prior distribution on the latent variable covariance
matrix.

While we prefer the
\code{"srs"} approach to covariances between observed variables, the
\code{"fa"} approach may be useful for complex models that run slowly. In these cases, it could be useful to sacrifice precise prior specification in favor of speed.

\subsection{Model fit and comparison} \label{sec:ppp}

Package \pkg{blavaan} supplies a variety of statistics for model
evaluation and comparison, available via the \code{fitMeasures()} function. This function is illustrated later in the examples, and specific statistics are described in Appendix~\ref{appx}. For model evaluation, \pkg{blavaan} supplies
posterior predictive checks of the model's log-likelihood
\citep[e.g.,][]{gelcarste04}.  For model comparision, it supplies a variety of statistics
including the 
Deviance Information Criterion \citep[DIC;][]{spibes02}, 
the (Laplace-approximated) log-Bayes factor, the Widely Applicable Information Criterion \citep[WAIC;][]{wat10},
and the leave-one-out
cross-validation statistic \citep[LOO; e.g.,][]{gel96}.
The latter two statistics are computed by \proglang{R} package
\pkg{loo} \citep{loo},
using output from \pkg{blavaan}.

Calculation of the information criteria is somewhat complicated by the fact that,
during \proglang{JAGS} model estimation, we condition on the 
latent variables $\bm{\eta}_i$, $i=1,\ldots,n$.  That is, our model utilizes likelihoods of the form $L(\bm{\vartheta},
\bm{\eta}_i | \bm{y}_i)$, where $\bm{\vartheta}$ is the vector of model parameters
and $\bm{\eta}_i$ is a vector of latent variables associated with 
individual $i$.  Because the latent variables are random effects,
we must integrate them out to obtain $L(\bm{\vartheta} |
\bm{Y})$, the likelihood by which model
assessment statistics are typically computed.  This is easy to do for the models considered
here, because the integrated likelihood continues to be multivariate
normal.  However, we cannot rely on \proglang{JAGS} to automatically
calculate the correct likelihood, so that \pkg{blavaan} calculates the
likelihood separately after parameters have been sampled in \proglang{JAGS}.

Now that we have provided background information on the models implemented in \pkg{blavaan}, the next section further describes how the package works. We then provide a series of examples.

% loo: cross validation and waic

\section[Overview of blavaan]{Overview of \pkg{blavaan}}

Readers familiar with \pkg{lavaan}  will also be
familiar with \pkg{blavaan}. The main functions are the same as the main \pkg{lavaan} functions, except that they start with the letter `b'.  For example, confirmatory factor analysis models are estimated via \code{bcfa()} and structural equation models are estimated via \code{bsem()}.  These two functions call the more general \code{blavaan()} function with a specific arrangement of arguments. The \pkg{blavaan} model specification syntax is nearly the same as the \pkg{lavaan} model specification syntax, and many functions are used by both packages.

%% Package \pkg{blavaan} is designed to estimate models as
%% quickly and efficiently as \proglang{JAGS} will allow.  The package makes use of
%% existing ``tricks'' for model estimation,
%% primarily replicating the \cite{lee07} strategy for SEM estimation in
%% \proglang{WinBUGS}.  In addition to this strategy, the parameter expansion approach outlined above 
%% allows us to estimate models that were difficult to estimate under Lee's
%% strategy.
% Model estimation via \proglang{JAGS} will never be the
% fastest possible, however.  For example, Bayesian estimation in \proglang{Mplus}
% is at least one order of magnitude faster than is estimation in
% \pkg{blavaan}, with \proglang{Mplus} taking seconds for models where
% \pkg{blavaan} may take 1--2 minutes.  This reflects an inherent trade-off
% between speed and flexibility (in, e.g., prior distribution
% specification and model extension).

As compared to \pkg{lavaan}, there are a small number of new features in 
\pkg{blavaan} that are specific to Bayesian modeling: prior distribution specification, export/use of 
\proglang{JAGS} syntax, convergence diagnostics, and specification of initial values.  We discuss these
topics in the context of a simple, one-factor confirmatory model.  The model uses the \cite{holswi39} data included with \pkg{lavaan}, with code for model specification and estimation appearing below.

\begin{Schunk}
\begin{Sinput}
> model <- ' visual =~ x1 + x2 + x3 '
> fit <- bcfa(model, data = HolzingerSwineford1939, jagfile = TRUE)
\end{Sinput}
\end{Schunk}

\subsection{Prior distribution specification}

Package \pkg{blavaan} defines a default prior distribution for each type of model parameter (see Equations~\eqref{eq:sem1} and~\eqref{eq:sem2}).  These default priors can be viewed via

\begin{Schunk}
\begin{Sinput}
> dpriors()
\end{Sinput}
\begin{Soutput}
             nu           alpha          lambda            beta          itheta 
"dnorm(0,1e-3)" "dnorm(0,1e-2)" "dnorm(0,1e-2)" "dnorm(0,1e-2)"  "dgamma(1,.5)" 
           ipsi             rho           ibpsi 
 "dgamma(1,.5)"    "dbeta(1,1)" "dwish(iden,3)" 
\end{Soutput}
\end{Schunk}

which includes a separate default prior for eight types of parameters.  Default prior distributions are placed on precisions instead of variances, so that the letter i in \code{itheta} and \code{ipsi} stands for ``inverse.''  Most of these parameters correspond to notation from Equations~\eqref{eq:sem1} to~\eqref{eq:lres}, with the exception of \code{rho} and \code{ibpsi}. The \code{rho} prior is used only for the \code{"srs"} option, as it is for correlation parameters associated with covariances in $\Theta$ or $\Psi$. The \code{ibpsi} prior is used for the covariance matrix of latent variables that are all allowed to covary with one another (commonly, blocks of exogenous latent variables). This block prior can improve sampling efficiency and reduce autocorrelation.

For most parameters, the user is free to declare prior distributions using any prior distribution that is defined in \proglang{JAGS}.  Changes to default priors can be supplied with the \code{dp} argument.  Further, the modifiers \code{[sd]} and \code{[var]} can be used to put a prior on a standard deviation or variance parameter, instead of the corresponding precision parameter.  For example,

\begin{Schunk}
\begin{Sinput}
> fit <- bcfa(model, data = HolzingerSwineford1939, 
+             dp = dpriors(nu = "dnorm(4,1)", itheta = "dunif(0,20)[sd]"))
\end{Sinput}
\end{Schunk}

sets the default priors for (i) the manifest variables' intercepts to normal with mean 4 and precision 1, and (ii) the manifest variables' standard deviations to uniform with bounds of 0 and 20.

Priors associated with the \code{rho} and \code{ibpsi} parameters are less amenable to change.  The default \code{rho} prior is a beta(1,1) distribution with support on $(-1,1)$; beta distributions with other parameter values could be used, but this distribution must be beta for now.  The default prior on blocks of precision parameters is the Wishart with identity scale matrix and degrees of freedom equal to the dimension of the scale matrix plus one.  The user cannot easily make changes to this prior distribution via the \code{dp} argument.  Changes can be made, however, by exporting the \proglang{JAGS} syntax and manually editing it.  This is further described in the next section.

In addition to priors on classes of model parameters, the user may wish to set the prior of a specific model parameter.  This is accomplished by using the \code{prior()} modifier within the model specification.  For example,

\begin{Schunk}
\begin{Sinput}
> model <- ' visual =~ x1 + prior("dnorm(1,1)")*x2 + x3 '
\end{Sinput}
\end{Schunk}

sets a specific prior distribution for the loading parameter from the visual factor to x2. Priors set in this manner (via the model syntax) take precedent over the default priors set via the \code{dp} argument.  Additionally, if one specifies a covariance parameter in the model syntax, the \code{prior()} modifier is for the correlation associated with the covariance, as opposed to the covariance itself.  The prior should be a distribution with support on $(0,1)$ (typically the beta distribution), which is automatically converted to an analogous distribution that has support on $(-1,1)$.

\subsection[JAGS syntax]{\proglang{JAGS} syntax}

Users may find that some desired features are not currently implemented in \pkg{blavaan}; such features may be related to, e.g., the use of specific types of priors or the handling of discrete variables. In these situations, we allow the user to export the \proglang{JAGS} syntax so that they can implement new features themselves.  We believe this will be especially useful for researchers who wish to develop new types of models: these researchers can specify a traditional model in \pkg{blavaan} that is related to the one that they want to implement, and they can then obtain the \proglang{JAGS} syntax and data for the traditional model.  This \proglang{JAGS} syntax should ease implementation of the novel model.

To export the \proglang{JAGS} syntax, users can employ the \code{jagfile} argument that was used in the \code{bcfa} command above. When \code{jagfile} is set to \code{TRUE}, the syntax will be written to the \code{lavExport} folder; if a character string is supplied, \pkg{blavaan} will use the character string as the folder name.

When the syntax is exported, two files are written within the folder.
The first, \code{sem.jag}, contains the model specification in
\pkg{JAGS} syntax.  The second, \code{semjags.rda}, is a list named
\code{jagtrans} that contains the data and initial values in
\proglang{JAGS} format, as well as labels associated with model
parameters.  These pieces can be used to run the model ``manually''
via, e.g., package \pkg{rjags} \citep{plu14} or \pkg{runjags} \citep{den15}.  The example below illustrates how to load the \proglang{JAGS} data (along with initial values and parameter labels) and estimate the exported model via \pkg{runjags}.

\begin{Schunk}
\begin{Sinput}
> load("lavExport/semjags.rda")
> fit <- run.jags("lavExport/sem.jag", monitor = jagtrans$coefvec$jlabel,
+                 data = jagtrans$data, inits = jagtrans$inits)
\end{Sinput}
\end{Schunk}

If the user modifies the \proglang{JAGS} file, then the \code{monitor} and \code{inits} arguments might require modification (or, for automatic initial values from \proglang{JAGS}, the user could omit the \code{inits} argument).  The data should not require modification unless the user adds or removes observed variables from the model.

We also note that the dimensions of the parameter vectors/matrices in the \proglang{JAGS} syntax generally match the dimensions that we would expect from Equations~\eqref{eq:sem1} and~\eqref{eq:sem2}, with a third dimension being added for multiple group models. These matrix dimensions (and the specific nonzero entries within each matrix) are obtained directly from \pkg{lavaan}.

\subsection{Convergence diagnostics}

Package \pkg{blavaan} offers two methods for monitoring chain convergence, specified as the \code{"auto"} or \code{"manual"} options to the \code{convergence} argument. Regardless of which option is used, the default number of chains is three and can be modified via the \code{n.chains} argument.

Under \code{"auto"} convergence, chains are sampled for an initial period in order to (i) achieve convergence as determined by the Potential Scale Reduction Factor \citep[PSRF;][]{gelrub92} 
and (ii) determine the number of samples necessary to obtain precise posterior estimates. Following this initial period, the chains are further sampled for the determined number of iterations. This automatic assessment comes directly from the \code{autorun.jags()} function of package \pkg{runjags} \citep{den15}; see there for further detail.

Under \code{"manual"} convergence, the user can specify the desired number of adaptation, burnin, and sample iterations via arguments of the same name (with defaults of 1,000 adaptation iterations, 4,000 burnin iterations, and 10,000 sample iterations). Adaptation iterations are those where samplers can modify themselves to increase sampling efficiency, whereas burnin iterations are discarded samples obtained from ``fixed'' samplers \citep[e.g.,][]{jags15}. A warning is issued if any PSRF is greater than 1.2, though some users may desire a more stringent criterion that is closer to 1.0. Access to the parameters' PSRF values is obtained via

\begin{Schunk}
\begin{Sinput}
> blavInspect(fit, "psrf")
\end{Sinput}
\end{Schunk}

Additionally, there is a \code{plot} method that includes the familiar time series plots, autocorrelation plots, and more. For example, autocorrelation plots for the first four free parameters are obtained via

\begin{Schunk}
\begin{Sinput}
> plot(fit, 1:4, "autocorr")
\end{Sinput}
\end{Schunk}

where parameter numbers correspond to the ordering of \code{coef(fit)}. This plot functionality comes from package \pkg{runjags}, with plotting options being found in the help file for \code{plot.runjags()}.

\subsection{Initial values}
In many Bayesian SEMs, poorly-chosen initial values can lead to extremely slow convergence. Package \pkg{blavaan} provides multiple options for supplying initial values that vary from chain to chain and/or that improve chain convergence. These are obtained via the \code{inits} argument. Under option \code{"prior"} (default), starting values are random draws from each parameter's prior distribution. However, to aid in convergence, loading and regression parameters are all required to be positive and close to 1, and correlation parameters are required to be close to zero. Under option \code{"jags"}, starting values are automatically set by \proglang{JAGS}.

Additional options for starting values are obtained as byproducts of \pkg{lavaan}. For example, the options \code{"Mplus"} and \code{"simple"} are analogous to the options for the \pkg{lavaan} \code{start} argument, where the same initial values are used for all chains. Individual initial values can also be set via the \code{start()} modifier to the lavaan syntax, and these initial values will be used for all chains.

Finally, it is also possible to supply a full set of user-defined starting values for each chain. This is most easily accomplished by estimating the model once in order to obtain the list of initial values. The initial values of a fitted model can be obtained via \code{blavInspect()}; i.e.,

\begin{Schunk}
\begin{Sinput}
> myinits <- blavInspect(fit, "inits")
\end{Sinput}
\end{Schunk}

The object \code{myinits} is a list whose length equals \code{n.chains}, with each entry being another list that contains initial values for the model's parameter vector. It may be helpful to call \code{blavInspect()} with the \code{"jagnames"} argument, in order to examine correspondence between \pkg{blavaan} parameter names and \proglang{JAGS} parameter names. Finally, after editing, a call to \pkg{blavaan} with the argument \code{inits = myinits} will estimate a model using the custom starting values.

Now that we have seen some features that are novel to \pkg{blavaan}, we will further illustrate the package by example.

\section{Applications}

In the applications below, we first illustrate some general features involving prior distribution specification and model fit measures.  We then illustrate the study of measurement invariance via Bayesian models, which involves multiple across-group parameter constraints. Finally, we provide some advanced examples involving the direct modification of exported \proglang{JAGS} code and the use of informative prior distributions.

\subsection{Political democracy}

We begin with the \cite{bol89} industrialization-democracy model discussed earlier in the paper.  The model specification is identical to the specification that we would use in \pkg{lavaan}:

\begin{Schunk}
\begin{Sinput}
> model <- ' 
+        # latent variable definitions
+          ind60 =~ x1 + x2 + x3
+          dem60 =~ y1 + a*y2 + b*y3 + c*y4
+          dem65 =~ y5 + a*y6 + b*y7 + c*y8
+      
+        # regressions
+          dem60 ~ ind60
+          dem65 ~ ind60 + dem60
+      
+        # residual correlations
+          y1 ~~ y5
+          y2 ~~ y4 + y6
+          y3 ~~ y7
+          y4 ~~ y8
+          y6 ~~ y8
+ '
\end{Sinput}
\end{Schunk}

though we could also use the \code{prior()} modifier to include custom priors within the model specification.  Instead, we specify priors for classes of parameters within the \code{bsem()} command below.  We first place $N(5,\sigma^{-2}=.01)$ priors on the observed-variable intercepts ($\nu$).  Next, following \cite{barmcc00}, we place log-normal priors on the residual standard deviations of both observed and latent variables.  This is accomplished by adding \code{[sd]} to the end of the \code{itheta} and \code{ipsi} priors.  \cite{barmcc00} stated that the log-normal priors result in better computational stability, as compared to the use of gamma priors on the precision parameters. 
%This option is not currently available in Mplus; there, we are restricted to gamma priors on the precisions that are related to the Wishart distribution.
Finally, we set beta(3,3) priors on the correlations associated with the residual covariance parameters.  This reflects the belief that the correlations will generally be closer to 0 than to $-1$ or $1$.  The result is the following command:

\begin{figure}
\begin{Schunk}

{\centering \includegraphics[width=4.5in,height=4.5in]{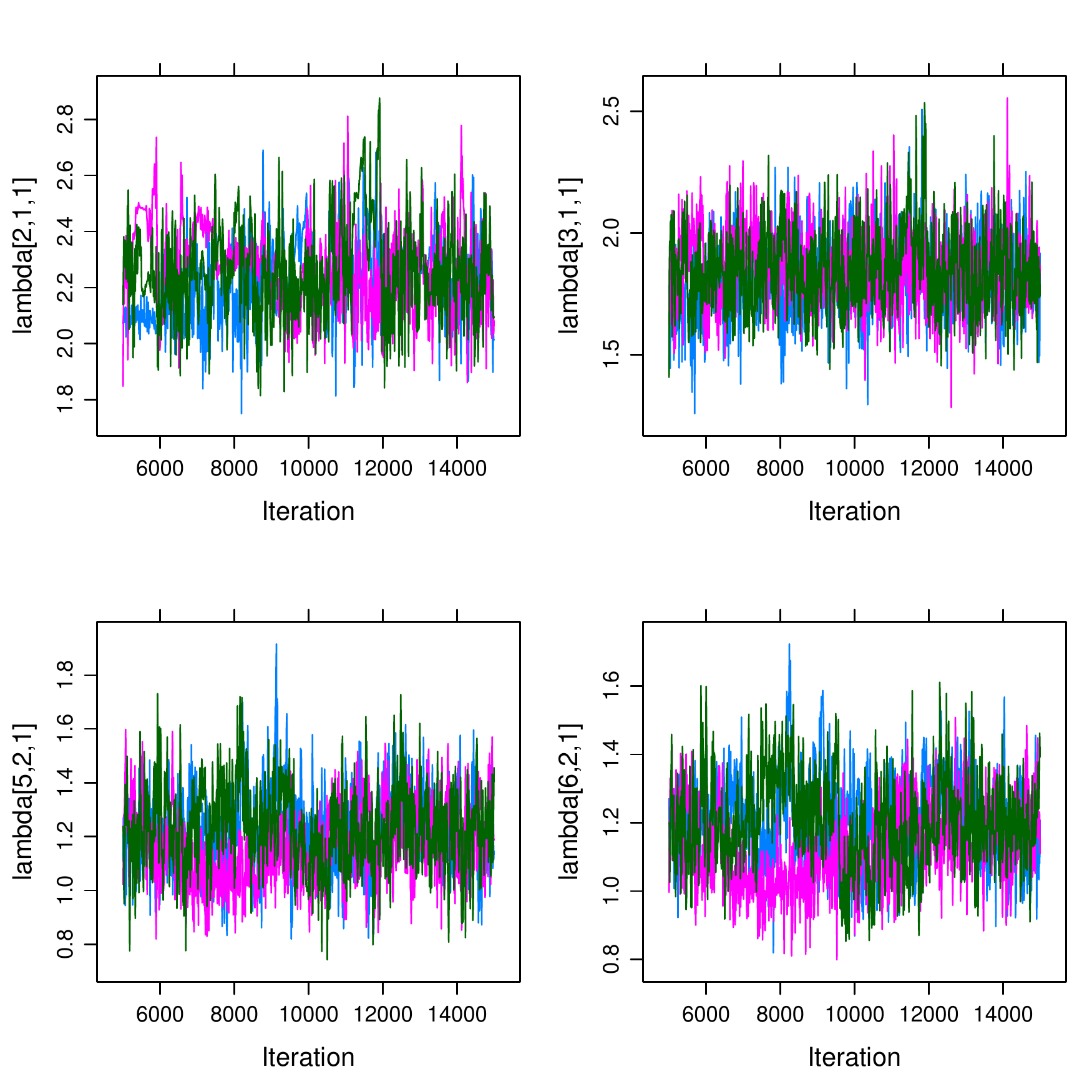} 

}

\end{Schunk}
\caption{Trace plots of the first four parameters from the political democracy model.}
\label{fig:tspl}
\end{figure}

\begin{Schunk}
\begin{Sinput}
> fit <- bsem(model, data = PoliticalDemocracy, 
+             dp = dpriors(nu = "dnorm(5,1e-2)", itheta = "dlnorm(1,.1)[sd]",
+                          ipsi = "dlnorm(1,.1)[sd]", rho = "dbeta(3,3)"),
+             jagcontrol = list(method = "rjparallel"))
\end{Sinput}
\end{Schunk}

where the latter argument, \code{jagcontrol}, allows us to sample each chain in parallel for faster model estimation.  Defaults for the number of adaptation, burnin, and drawn samples come from package \pkg{runjags}; those defaults are currently 1,000, 4,000, and 10,000, respectively. Users can modify any of these via the arguments \code{adapt}, \code{burnin}, and \code{sample}. Trace plots of sampled parameters can be obtained immediately; for example, the command below provides trace plots for the first four free parameters, with the resulting plot displayed in Figure~\ref{fig:tspl}.

\begin{Schunk}
\begin{Sinput}
> plot(fit, 1:4, "trace")
\end{Sinput}
\end{Schunk}

As described earlier in the paper, users can also control the handling of residual covariances via the \code{cp} argument. For the particular model considered here, we compared the sampling speed and efficiency (excluding posterior predictive checks) of the \code{"srs"} and \code{"fa"} options using a four-core Dell Precision laptop running Ubuntu linux. We observed that the \code{"srs"} option took 76 seconds and the \code{"fa"} option took 65 seconds, illustrating the speed advantage of the \code{"fa"} approach. However, the ratio of effective sample sizes for the two approaches averaged 1.44 in favor of the \code{"srs"} approach (where the average is taken across parameters), implying that \code{"srs"} leads to more efficient sampling. In our experience, these results can vary depending on the specific model of interest, but it generally appears that \code{"fa"} is somewhat faster and \code{"srs"} somewhat more efficient.

% speed test detail here

% summary()
Once we have sampled for the desired number of iterations, we can make use of \pkg{lavaan} functions in order to summarize and study the fitted model.  Users may primarily be interested in \code{summary()}, which organizes results in a manner similar to the classical \pkg{lavaan} models.

% reduced font size so the output displays nicely
{\footnotesize
\begin{Schunk}
\begin{Sinput}
> summary(fit)
\end{Sinput}
\begin{Soutput}
blavaan (0.2-2) results of 10000 samples after 5000 adapt+burnin iterations

  Number of observations                            75

  Number of missing patterns                         1

  Statistic                                 MargLogLik         PPP
  Value                                      -1659.057       0.556

Parameter Estimates:


Latent Variables:
                   Estimate  Post.SD  HPD.025  HPD.975     PSRF    Prior        
  ind60 =~                                                                      
    x1                1.000                                                     
    x2                2.249    0.152    1.957    2.544    1.018    dnorm(0,1e-2)
    x3                1.839    0.167    1.517    2.169    1.016    dnorm(0,1e-2)
  dem60 =~                                                                      
    y1                1.000                                                     
    y2         (a)    1.208    0.152    0.922    1.513    1.055    dnorm(0,1e-2)
    y3         (b)    1.177    0.138     0.92    1.448    1.103    dnorm(0,1e-2)
    y4         (c)    1.261    0.142     0.99     1.54    1.108    dnorm(0,1e-2)
  dem65 =~                                                                      
    y5                1.000                                                     
    y6         (a)    1.208    0.152    0.922    1.513    1.055                 
    y7         (b)    1.177    0.138     0.92    1.448    1.103                 
    y8         (c)    1.261    0.142     0.99     1.54    1.108                 

Regressions:
                   Estimate  Post.SD  HPD.025  HPD.975     PSRF    Prior        
  dem60 ~                                                                       
    ind60             1.426    0.415    0.618    2.246    1.011    dnorm(0,1e-2)
  dem65 ~                                                                       
    ind60             0.616    0.260    0.095    1.029    1.038    dnorm(0,1e-2)
    dem60             0.881    0.077    0.747    1.041    1.028    dnorm(0,1e-2)

Covariances:
                   Estimate  Post.SD  HPD.025  HPD.975     PSRF    Prior        
 .y1 ~~                                                                         
   .y5                0.544    0.391   -0.201    1.302    1.059       dbeta(3,3)
 .y2 ~~                                                                         
   .y4                1.459    0.702    0.144    2.919    1.009       dbeta(3,3)
   .y6                2.117    0.752    0.684    3.608    1.000       dbeta(3,3)
 .y3 ~~                                                                         
   .y7                0.744    0.655   -0.547    2.015    1.019       dbeta(3,3)
 .y4 ~~                                                                         
   .y8                0.392    0.494   -0.515    1.413    1.024       dbeta(3,3)
 .y6 ~~                                                                         
   .y8                1.357    0.573    0.249    2.491    1.011       dbeta(3,3)

Intercepts:
                   Estimate  Post.SD  HPD.025  HPD.975     PSRF    Prior        
   .x1                5.045    0.080    4.895    5.202    1.016    dnorm(5,1e-2)
   .x2                4.772    0.163    4.449    5.078    1.020    dnorm(5,1e-2)
   .x3                3.543    0.157     3.24    3.847    1.013    dnorm(5,1e-2)
   .y1                5.475    0.276    4.925     6.03    1.057    dnorm(5,1e-2)
   .y2                4.269    0.425    3.427    5.097    1.037    dnorm(5,1e-2)
   .y3                6.571    0.377    5.821    7.295    1.041    dnorm(5,1e-2)
   .y4                4.464    0.361    3.748    5.164    1.049    dnorm(5,1e-2)
   .y5                5.142    0.288    4.567    5.699    1.050    dnorm(5,1e-2)
   .y6                2.981    0.377     2.22    3.706    1.042    dnorm(5,1e-2)
   .y7                6.201    0.348    5.507    6.879    1.044    dnorm(5,1e-2)
   .y8                4.047    0.357    3.371    4.773    1.053    dnorm(5,1e-2)
    ind60             0.000                                                     
   .dem60             0.000                                                     
   .dem65             0.000                                                     

Variances:
                   Estimate  Post.SD  HPD.025  HPD.975     PSRF    Prior        
   .x1                0.096    0.023    0.051    0.139    1.003 dlnorm(1,.1)[sd]
   .x2                0.086    0.080        0     0.24    1.008 dlnorm(1,.1)[sd]
   .x3                0.524    0.101    0.334    0.723    1.002 dlnorm(1,.1)[sd]
   .y1                1.933    0.577    0.774    3.012    1.164 dlnorm(1,.1)[sd]
   .y2                7.959    1.444    5.293    10.84    1.002 dlnorm(1,.1)[sd]
   .y3                5.376    1.064    3.445    7.496    1.011 dlnorm(1,.1)[sd]
   .y4                3.552    0.870    1.911     5.33    1.035 dlnorm(1,.1)[sd]
   .y5                2.507    0.518    1.553    3.556    1.012 dlnorm(1,.1)[sd]
   .y6                5.224    0.962    3.477     7.21    1.005 dlnorm(1,.1)[sd]
   .y7                3.902    0.819     2.35    5.487    1.018 dlnorm(1,.1)[sd]
   .y8                3.559    0.765    2.086    5.047    1.019 dlnorm(1,.1)[sd]
    ind60             0.451    0.092    0.287    0.639    1.008 dlnorm(1,.1)[sd]
   .dem60             4.088    1.025    2.199    6.105    1.057 dlnorm(1,.1)[sd]
   .dem65             0.117    0.165        0    0.468    1.156 dlnorm(1,.1)[sd]
\end{Soutput}
\end{Schunk}
}

While the results look similar to those that one would see in \pkg{lavaan}, there are a few differences that require explanation.  First, the top of the output includes two model evaluation measures: a Laplace approximation of the marginal log-likelihood and the posterior predictive $p$-value (see Appendix~\ref{appx}).  Second, the ``Parameter Estimates'' section contains many new columns.  These are the posterior mean (\code{Post.Mean}), the posterior standard deviation (\code{Post.SD}), a 95\% highest posterior density interval (\code{HPD.025} and \code{HPD.975}), the potential scale reduction factor for assessing chain convergence \citep[\code{PSRF};][]{gelrub92}, and the prior distribution associated with each model parameter (\code{Prior}). Users can optionally obtain posterior medians, posterior modes, and effective sample sizes (number of posterior samples drawn, adjusted for autocorrelation).  These can be obtained by supplying the logical arguments \code{postmedian}, \code{postmode}, and \code{neff} to \code{summary()}.

Finally, we can obtain various Bayesian model assessment measures via \code{fitMeasures()}

\begin{Schunk}
\begin{Sinput}
> fitMeasures(fit)
\end{Sinput}
\begin{Soutput}
      npar       logl        ppp        bic        dic      p_dic       waic 
    39.000  -1550.336      0.556   3268.531   3174.007     36.667   3178.144 
    p_waic      looic      p_loo margloglik 
    38.184   3178.744     38.484  -1659.057 
\end{Soutput}
\end{Schunk}

where, as previously mentioned, the WAIC and LOO statistics are computed with the help of package \pkg{loo}.  Other \pkg{lavaan} functions, including \code{parameterEstimates()} and \code{parTable()}, similarly work as expected.

\subsection{Measurement invariance}

Verhagen and Fox \citep[\citeyear{verfox13}; see also][]{verlev15}
describe the use of Bayesian methods for studying
the measurement invariance of sets of items or scales.  Briefly,
measurement invariance means that the items or scales measure diverse
groups of individuals in a fair manner.  For example, two individuals with the same
level of mathematics ability should receive the same score on a
mathematics test (within random
noise), regardless of the individuals' backgrounds and demographic
variables.  In this section, we illustrate a Bayesian measurement
invariance approach using the \cite{holswi39} data.  This section also
illustrates how we can estimate a model in \pkg{blavaan} using pieces of a fitted \pkg{lavaan} object. This may be of interest to advanced users who would prefer to write/edit a lavaan ``parameter table'' instead of using model specification syntax. Additionally, the use of a \pkg{lavaan} object provides a path from \proglang{Mplus} to \pkg{blavaan}, via the \code{mplus2lavaan()} function from \pkg{lavaan}.

We first fit three increasingly-restricted models to the
\cite{holswi39} data via classical methods.  This is accomplished via
the \pkg{lavaan} \code{cfa()} function (though we could also
immediately use \code{bcfa()} here):

\begin{Schunk}
\begin{Sinput}
> HS.model <- ' visual  =~ x1 + x2 + x3
+               textual =~ x4 + x5 + x6
+               speed   =~ x7 + x8 + x9 '
> fit1 <- cfa(HS.model, data = HolzingerSwineford1939, group = "school")
> fit2 <- cfa(HS.model, data = HolzingerSwineford1939, group = "school",
+              group.equal = "loadings")
> fit3 <- cfa(HS.model, data = HolzingerSwineford1939, group = "school",
+              group.equal = c("loadings", "intercepts"))
\end{Sinput}
\end{Schunk}

The resulting objects each include a ``parameter table,'' which contains all model details necessary for \pkg{lavaan} estimation.  To fit these models via Bayesian methods, we can merely pass the classical model's parameter table and data to \pkg{blavaan}.  This is accomplished via

\begin{Schunk}
\begin{Sinput}
> bfit1 <- bcfa(parTable(fit1), data = HolzingerSwineford1939, 
+               group = "school")
> bfit2 <- bcfa(parTable(fit2), data = HolzingerSwineford1939, 
+               group = "school")
> bfit3 <- bcfa(parTable(fit3), data = HolzingerSwineford1939, 
+               group = "school")
\end{Sinput}
\end{Schunk}

As mentioned above, we could also supply the \code{group.equal} argument directly to the \code{bcfa()} calls.

Following model estimation, we can immediately compare the models via \code{fitMeasures()}.  For example, focusing on the first two models, we obtain

\begin{Schunk}
\begin{Sinput}
> fitMeasures(bfit1)
\end{Sinput}
\begin{Soutput}
      npar       logl        ppp        bic        dic      p_dic       waic 
    60.000  -3684.433      0.000   7710.893   7482.783     56.958   7492.325 
    p_waic      looic      p_loo margloglik 
    64.291   7492.597     64.427  -3937.715 
\end{Soutput}
\begin{Sinput}
> fitMeasures(bfit2)
\end{Sinput}
\begin{Soutput}
      npar       logl        ppp        bic        dic      p_dic       waic 
    54.000  -3687.266      0.000   7682.356   7478.285     51.877   7485.299 
    p_waic      looic      p_loo margloglik 
    56.939   7485.548     57.063  -3917.929 
\end{Soutput}
\end{Schunk}

As judged by the posterior predictive $p$-value (\code{ppp}), neither of the two models provide a good fit to the data.  In practice, we might stop our measurement invariance study there.  However, for the purpose of demonstration, we can also examine the information criteria.  Based on these, we would prefer the second model (due to the smaller DIC, WAIC, and LOOIC values).  %% In addition to these measures, we have implemented a Laplace approximation to the log-Bayes factor in the \code{BF()} function.  The Bayes factor results are currently labelled as experimental: while we have not found a reason to doubt the results, we also do not yet fully understand the results.  We plan further study of these issues.

\subsection{Informative prior distributions}
We have seen that \pkg{blavaan} allows users to specify prior distributions for individual model parameters and for classes of model parameters. In this section, we illustrate how informative prior distributions can be set for factor analysis parameters in the context of a specific dataset, where priors are set based on the parameters' interpretations.

We consider a one-factor, three-indicator model fit to a subset of data from a stereotype threat study \citep{wicdol05}. The data are available in \proglang{R} package \pkg{psychotools} \citep{zeistr15}, and the model specifies a general ``ability'' factor underlying three academic tests (of verbal ability, numerical ability, and abstract reasoning). The test scores are sums of the number of items answered correctly, so they are bounded from below at 0 and from above at 16 (verbal), 14 (numerical), and 18 (abstract reasoning). While these bounds technically violate the normality assumption of the factor analysis model, it is customary to apply this model to test scores (especially in the current situation, where the test scores are unimodal with modes in the middle of the score ranges).
A factor analysis model with default (non-informative) prior distributions and automatic convergence assessment can be fit to the data via

\begin{Schunk}
\begin{Sinput}
> library("psychotools")
> data("StereotypeThreat")
> st <- subset(StereotypeThreat, ethnicity == "majority")
> model <- 'ability =~ abstract + verbal + numerical'
> bfit <- bcfa(model, data = st, convergence = "auto")
\end{Sinput}
\end{Schunk}

Instead of using the default prior distributions, we can tailor prior distributions to this particular dataset by simultaneously considering the data and each model parameter's interpretation. First, because the test scores are all bounded, we know the maximum possible standard deviation on each test (obtained if half the participants scored a 0 and half score the maximum). These are 8 for the verbal test, 7 for the numerical test, and 9 for the abstract reasoning test. Thus, we place uniform priors on the manifest variables' standard deviations with lower bounds at 0 and upper bounds at these maxima:

\begin{Schunk}
\begin{Sinput}
> model <- c(model,
+            'abstract ~~ prior("dunif(0,9)[sd]") * abstract
+             verbal ~~ prior("dunif(0,8)[sd]") * verbal
+             numerical ~~ prior("dunif(0,7)[sd]") * numerical')
\end{Sinput}
\end{Schunk}

Next, we consider the factor variance. Because the loading associated with the abstract reasoning test will be fixed at 1, the factor variance can be interpreted as the variability in the abstract reasoning test that can be attributed to the ability factor. Because the three tests were chosen to be highly correlated (and because the total standard deviation of abstract reasoning is 9 or less), we use a uniform prior with an upper bound of 6 here. Thus, even in the case where we observe the maximum standard deviation of 9, the factor could still account for $2/3$ of the variability in the ability test.

\begin{Schunk}
\begin{Sinput}
> model <- c(model,
+            'ability ~~ prior("dunif(0,4.5)[sd]") * ability')
\end{Sinput}
\end{Schunk}

Next, we consider the manifest variables' intercept parameters, which reflect average scores on each of the three tests. These tests are known historically to result in average scores near the middle of the range \citep{wicdol05}, so our prior distributions are truncated normals centered at the middle of each test's range (with the truncation points at the minimum and maximum possible scores on each test):

\begin{Schunk}
\begin{Sinput}
> model <- c(model,
+            'abstract ~ prior("dnorm(9,.25)T(0,18)") * 1
+             verbal ~ prior("dnorm(8,.25)T(0,16)") * 1
+             numerical ~ prior("dnorm(7,.25)T(0,14)") * 1')
\end{Sinput}
\end{Schunk}

Finally, we consider the factor loadings. We have no knowledge that the ability factor will influence one test more than others, so we might expect the two free loadings to be around 1 (because the fixed loading equals 1). We also expect positive loadings, because ability should positively influence all three tests. Thus, we supply uniform(0,3) priors to the \code{dp} argument of \code{bcfa()}:

\begin{Schunk}
\begin{Sinput}
> bfitInform <- bcfa(model, data = st,
+                    dp = dpriors(lambda = "dunif(0,3)"),
+                    convergence = "auto")
\end{Sinput}
\end{Schunk}

The two models' fit measures are generally similar and not shown. We compare the two models' posterior means and standard deviations in Table~\ref{tab:wicpri}. The differences are unlikely to impact substantive conclusions, but two of them are noteworthy. First, the factor variance $\psi_1$ is larger under the model with informative priors, likely because the informative prior (uniform(0,4.5) on the standard deviation) placed more density on larger values of the standard deviation. We observe a similar phenomenon with the residual variance of the numerical test ($\theta_3$). Second, the posterior means and standard deviations of the loadings ($\lambda_2$ and $\lambda_3$) are somewhat smaller under the informative priors. This is likely related to the larger variance estimates.

\begin{table}
\caption{Comparison of posterior means and standard deviations under default prior distributions and informative prior distributions.}
\label{tab:wicpri}
\begin{center}
\begin{tabular}{llrrrrrrrrr}
    \hline
    & & \multicolumn{9}{c}{Parameter} \\
Prior &  & $\lambda_2$ & $\lambda_3$ & $\theta_1$ & $\theta_2$ & $\theta_3$ & $\psi_1$ & $\nu_1$ & $\nu_2$ & $\nu_3$ \\ \hline
Default & Est & 1.14 & 1.66 & 7.96 & 8.45 & 1.43 & 1.93 & 9.84 & 6.96 & 5.43 \\
 & SD & 0.39 & 0.53 & 1.13 & 1.23 & 1.06 & 0.95 & 0.25 & 0.26 & 0.19 \\ \hline
Inform. & Est & 1.06 & 1.31 & 7.7 & 8.28 & 2.19 & 2.57 & 9.83 & 6.99 & 5.44 \\
 & SD & 0.33 & 0.43 & 1.21 & 1.28 & 1.1 & 1.13 & 0.25 & 0.26 & 0.19 \\ \hline

\end{tabular}
\end{center}
\end{table}

This example could be used as the start of a larger analysis of posterior sensitivity to prior distributions. For the purpose of automation, prior distributions could be entered directly into the model's parameter table (obtained via, e.g., \code{parTable(bfit)}) and the model subsequently re-estimated, similar to what was done in the measurement invariance example.

\subsection[Extensions of JAGS syntax]{Extensions of \proglang{JAGS} syntax}
Finally, we provide an example involving use of the JAGS syntax to estimate novel models. Consider the robust factor analysis model of \cite{zhali14}, which essentially replaces the factor analysis model's normal distributions with $t$ distributions. In particular, each factor $\eta$ arises from a $t_{\text{df}}(0,1)$ distribution, and each residual $\epsilon_j$ arises from a $t_{\text{df}}(0,\psi_j)$ distribution. The degrees of freedom, $\text{df}$, is a free parameter and is shared by all the $t$ distributions. To our knowledge, there exists no software to readily estimate this model, and \cite{zhali14} implemented their own Gibbs sampler (making use of analytic posterior distributions under conjugate priors). Here, we show how the \proglang{JAGS} syntax from \pkg{blavaan} can be modified to easily estimate this model. For illustration, we again apply a one-factor model to the \cite{wicdol05} data from the previous section.

We begin by using \pkg{blavaan} to export \proglang{JAGS} syntax for a simple, one-factor model.

\begin{Schunk}
\begin{Sinput}
> model <- 'ability =~ abstract + verbal + numerical'
> bfit <- bcfa(model, data = st, jagfile = TRUE)
\end{Sinput}
\end{Schunk}

\begin{figure}
\begin{Verbatim}[commandchars=\\\{\}]
model \{
  for(i in 1:N) \{
    abstract[i] ~ \textcolor{red}{dt}(mu[i,1], 1/theta[1,1,g[i]]\textcolor{red}{, df})
    verbal[i] ~ \textcolor{red}{dt}(mu[i,2], 1/theta[2,2,g[i]]\textcolor{red}{, df})
    numerical[i] ~ \textcolor{red}{dt}(mu[i,3], 1/theta[3,3,g[i]]\textcolor{red}{, df})

    mu[i,1] <- nu[1,1,g[i]] + lambda[1,1,g[i]]*eta[i,1]
    mu[i,2] <- nu[2,1,g[i]] + lambda[2,1,g[i]]*eta[i,1]
    mu[i,3] <- nu[3,1,g[i]] + lambda[3,1,g[i]]*eta[i,1]

    # lvs
    eta[i,1] ~ \textcolor{red}{dt}(mu.eta[i,1], 1/psi[1,1,g[i]]\textcolor{red}{, df})

    mu.eta[i,1] <- alpha[1,1,g[i]]
  \}

\textcolor{red}{  df <- 1/dfinv}
\textcolor{red}{  dfinv ~ dunif(1/200, 1)}

  # Assignments from parameter vector
\end{Verbatim}
\caption{Illustration of \proglang{JAGS} modifications necessary to implement the \cite{zhali14} robust factor analysis model. The original \proglang{JAGS} code exported from \pkg{blavaan} is in black font, with modifications and additions in red. Additional code requiring no modification (extending below the ``Assigments from parameter vector'' comment) is omitted.}
\label{fig:jagsyn}
\end{figure}

By default, the exported \proglang{JAGS} code is written to the file \code{lavExport/sem.jag}. Upon opening that file, we must make two edits (see Figure~\ref{fig:jagsyn}). First, the normal distributions (\code{dnorm()}) are replaced with $t$ distributions (\code{dt()}). Second, we need to specify a prior distribution for the degrees of freedom parameter. Similar to \cite{zhali14}, we place a flat prior on the inverse degrees of freedom. The lower bound is nonzero to aid in convergence; this is justified by realizing that, as the degrees of freedom increase, the $t$ distribution becomes the normal distribution. Once we surpass, say, 200 degrees of freedom, we can accept that larger values are practically equivalent to 200.

The modified syntax from Figure~\ref{fig:jagsyn} could then be estimated manually, making use of the exported data and adding a monitor for the \code{df} parameter. For example, if the exported files are saved as \code{lavExport/robustfa.jag} and \code{lavExport/robustfa.rda}, the model can be estimated via

\begin{Schunk}
\begin{Sinput}
> load("lavExport/robustfa.rda")
> fit <- run.jags("lavExport/robustfa.jag", 
+                 monitor = c(jagtrans$monitors, "df"),
+                 data = jagtrans$data, inits = jagtrans$inits)
\end{Sinput}
\end{Schunk}

While many model extensions will be more complicated than the one considered here, this example illustrates \pkg{blavaan}'s potential use for statistical researchers who are developing new models: instead of writing \proglang{JAGS} syntax entirely from scratch, these researchers may use \pkg{blavaan} to obtain syntax for a basic model that is similar to the desired model. In many cases, this will simplify implementation of the desired model.

\section{Conclusion}

As described throughout the paper, package \pkg{blavaan} combines many existing tools so that it is easy to estimate multivariate normal SEMs via open-source software. The package can be useful to applied researchers who need an expanded set of prior distributions at their disposal, freeing them from the need to learn the intricacies of an MCMC package. It can also be useful to methodological researchers, freeing them from the need to code their own \proglang{JAGS} models from scratch. The idea of separating model specification from MCMC coding seems generally useful beyond SEM, and others are indeed making progress in this direction. For example, package \pkg{rstanarm} \citep{rstanarm16} allows users to estimate generalized linear mixed models via \proglang{Stan}, using \pkg{lme4} \citep{lme415} syntax.
The coding of complex models in \proglang{BUGS}, \proglang{JAGS}, or \proglang{Stan} is often tedious even for experienced users, so that progress here could improve most users' workflows.

There are a variety of additional models that we plan to support in future versions of \pkg{blavaan}, including models with latent interactions, with ordinal variables, and with mixture and multilevel components. These models have been addressed in the literature \citep[e.g.,][]{aspmut10,sonlee01,sonlee12},
and \proglang{JAGS} examples for estimating specific instances of these models is available \citep[e.g.,][]{chopre15,merwan16}. Work remains, however, to sample from these models efficiently and to specify the models via \pkg{lavaan}'s syntax. The sampling efficiency of \proglang{Stan} may be useful for at least some of these models, and we plan to explore this in the future.\\ \ \\

%need?
%\subsection[Summary]{Summary}
In summary, \pkg{blavaan} is currently useful for estimating many types of multivariate normal SEMs, and the \proglang{JAGS} export feature allows researchers to extend the models in any fashion desired.  As additional features are added, we hope that the package keeps pace with \pkg{lavaan} as an open set of tools for SEM estimation and study.

\section{Acknowledgments}

This research was partially supported by NSF grants SES-1061334 and
1460719. The authors thank Herbert Hoijtink, Ross Jacobucci, two anonymous reviewers, and the editor for comments that helped to improve the paper. The authors are solely responsible for all remaining errors.

\bibliography{refs}

\appendix

\section{Details on model assessment statistics} \label{appx}

In the subsections below, we review the model evaluation and comparison metrics that \pkg{blavaan} supplies.

\subsubsection{Posterior predictive checks} 
As a measure of the model's absolute fit, \pkg{blavaan} computes a
posterior predictive $p$-value that compares observed likelihood ratio
test ($\chi^2$)
statistics to likelihood ratio test statistics generated from the model's posterior
predictive distribution \citep[also see][]{mutasp12,schhoi99}.  The likelihood ratio test statistic is generally
computed via
\begin{equation*}
    \text{LRT}(\bm{Y}, \bm{\vartheta}) = -2 \log L(\bm{\vartheta} | \bm{Y}) + 2 \log L(\bm{\vartheta}^{\text{sat}} | \bm{Y}),
\end{equation*}
where $\bm{\vartheta}^{\text{sat}}$ is a ``saturated'' parameter vector that
perfectly matches the observed mean and covariance matrix.

For each posterior draw $\bm{\vartheta}^s$ ($s=1,\ldots,S$), computation of
the posterior predictive $p$-value includes four steps:
\begin{enumerate}
  \item Compute the observed LRT statistic as $\text{LRT}(\bm{Y},
\bm{\vartheta}^s)$.
  \item Generate artificial data $\bm{Y}^{\text{rep}}$ from the model, assuming parameter
    values equal to $\bm{\vartheta}^s$.
  \item Compute the posterior predictive LRT statistic
    $\text{LRT}(\bm{Y}^{\text{rep}}, \bm{\vartheta}^s)$.
  \item Record whether or not $\text{LRT}(\bm{Y}^{\text{rep}},
    \bm{\vartheta}^s) > \text{LRT}(\bm{Y}, \bm{\vartheta}^s)$.
\end{enumerate}
The posterior predictive $p$-value is then the proportion of the time
(out of $S$ draws) that the posterior predictive LRT statistic is
larger than the observed LRT statistic.  Values closer to .5 indicate a
model that fits the observed data, while values closer to 0 indicate
the opposite. For use in practice, \cite{mutasp12} use a threshold of .05 (analogous to a frequentist $\alpha$ level) and present some evidence that, as compared to the frequentist likelihood ratio test, the posterior predictive $p$ is less sensitive to minor model misspecifications and exhibits better performance at small sample sizes. On the other hand, \cite{hoivan16} show that the posterior predictive $p$-value misbehaves in situations involving highly informative prior distributions (as are sometimes used for shrinkage/regularization). They recommend a prior-posterior predictive $p$-value that may be implemented in future versions of \pkg{blavaan}.

In \pkg{blavaan}, the posterior predictive $p$-value is currently impractical when there are missing data. This is 
because it is impractical to compute the saturated log-likelihood in the
presence of missing data.  We
need to use, e.g., the EM algorithm to do this, and this needs to be
done $(S+1)$ times (once for the saturated likelihood of the observed
data and once for the saturated likelihood of each of the $S$ artificial datasets).
With complete data, we can more simply calculate the saturated log-likelihood
using the sample mean vector and covariance matrix.  
% TODO add argument ppp="none"?
The argument
\code{test="none"} can be supplied to \pkg{blavaan} in order to
bypass these slow calculations when there are missing data, and future versions may sample the missing observations in order to avoid the issue.

% dic
\subsubsection{DIC}
The DIC is given as
\begin{equation}
  \label{eq:dic}
  DIC = -2 \log L(\hat{\bm{\vartheta}} | \bm{Y}) + 2\ \text{efp}_{\text{DIC}},
\end{equation}
where $\hat{\bm{\vartheta}}$ is a vector of posterior parameter means (or other measure of central tendency), $L(\hat{\bm{\vartheta}} | \bm{Y})$ is the model's likelihood at the posterior means, and $\text{efp}_{\text{DIC}}$ is the model's effective number of parameters, calculated as
\begin{equation}
  \label{eq:efp}
  \text{efp}_{\text{DIC}} = 2 \left [ \log L(\hat{\bm{\vartheta}} | \bm{Y}) - \overline{\log L(\bm{\vartheta} | \bm{Y})} \right ].
\end{equation}
The latter term, $\overline{\log L(\bm{\vartheta} | \bm{Y})}$, is obtained by calculating the log-likelihood for each posterior sample and then averaging.

Many readers will be familiar with the automatic calculation of DIC within
programs like \proglang{BUGS} and \proglang{JAGS}.  As we described
above, the automatic
DIC is not what users typically desire because the likelihood
conditions on the
latent variables $\bm{\eta}_i$.  This greatly
increases the effective number of parameters and may result in poor
inferences \citep[for further discussion of this issue, see][]{mil09}.
As previously mentioned, \pkg{blavaan} avoids 
the automatic DIC computation in \proglang{JAGS}, calculating its own
likelihoods after model parameters have been sampled.

\subsubsection{WAIC and LOO}
The WAIC and LOO are asymptotically equivalent measures that are
advantageous over DIC while also being more difficult to compute than
DIC.  Many computational difficulties are overcome in developments by
\cite{vehgel15}, with their methodology being implemented in package
\pkg{loo} \citep{loo}.  As input, \pkg{loo} requires casewise
log-likelihoods associated with a set of posterior samples: $\log L(\bm{\vartheta}^s | \bm{y}_i), s=1,\ldots,S; i=1,\ldots,n$.

Like DIC, both of these measures seek to characterize a model's predictive
accuracy (generalizability).  The definition of WAIC looks 
similar to that of DIC:
\begin{equation*}
    \text{WAIC} = -2 \text{lppd} + 2\ \text{efp}_{\text{WAIC}},
\end{equation*}
where the first term is related to log-likelihoods of observed data and the second term
involves an effective number of parameters.  Both terms now involve
log-likelihoods associated with individual observations, however, which help us estimate
the model's expected log pointwise predictive density (a measure of
predictive accuracy).  

The first term, the log pointwise predictive density of
the observed data (lppd), is estimated via
\begin{equation}
    \label{eq:lppd}
    \text{lppd} = \displaystyle \sum_{i=1}^n \log \left ( \frac{1}{S}
        \displaystyle \sum_{s=1}^S f(\bm{y}_i | \bm{\vartheta}^s) \right ),
\end{equation}
where $S$ is the number of posterior draws and $f(\bm{y}_i |
\bm{\vartheta}^s)$ is the density of observation $i$ with respect to the parameters
sampled at iteration $s$.  The second term, the effective number of
parameters, is estimated via
\begin{equation*}
    \text{efp}_{\text{WAIC}} = \displaystyle \sum_{i=1}^n \text{var}_s
    (\log f(\bm{y}_i | \bm{\vartheta})),
\end{equation*}
where we compute a separate variance for each observation $i$ across
the $S$ posterior draws.

The LOO measure estimates the predictive density of each
individual observation from a cross-validation standpoint; that is, the
predictive density when we hold out one observation at a time and use
the remaining observations to update the prior.  We can use analytical
results to estimate this quantity based on the full-data
posterior, removing the need to re-estimate the posterior while
sequentially holding out each observation.  These results lead to the estimate
\begin{equation}
    \label{eq:loo}
    \text{LOO} = -2 \displaystyle \sum_{i=1}^n \log \left (
        \frac{\sum_{s=1}^S w_i^s f(\bm{y}_i | \bm{\vartheta}^s)}{\sum_{s=1}^S
          w_i^s} \right ),
\end{equation}
where the $w_i^s$ are importance sampling weights based on the
relative magnitude of individual $i$'s density function across the $S$
posterior samples.  Package \pkg{loo} smooths these weights via a
generalized Pareto distribution, improving the estimate.

%% loo effective sample size
We can also estimate the effective number of parameters under the LOO
measure by comparing LOO to the lppd that was used in the WAIC
calculation.  This gives us
\begin{equation*}
    \text{efp}_{\text{LOO}} = \text{lppd} + \text{LOO}/2,
\end{equation*}
where these terms come from Equations~\eqref{eq:lppd}
and~\eqref{eq:loo}, respectively.  Division of the latter term by two
(and addition, vs.\ subtraction) offsets the multiplication by $-2$ that
occurs in~\eqref{eq:loo}.  
For further detail on all these measures, see \cite{vehgel15}.

\subsubsection{Bayes factor}
% laplace approx to bf
The Bayes factor between two models is defined as the ratio of the models' marginal
likelihoods \citep[e.g.,][]{kasraf95}.  Candidate model 1's marginal
likelihood can be written as 
\begin{equation}
    \label{eq:ml}
    f_1(\bm{Y}) = \displaystyle \int f(\bm{\vartheta}_1) f(\bm{Y} |
    \bm{\vartheta}_1) \partial \bm{\vartheta}_1,
\end{equation}
where $\bm{\vartheta}_1$ is a vector containing candidate model 1's free
parameters (excluding the latent variables $\bm{\eta}_i$) and $f()$ is
a probability density function.  The marginal
likelihood of candidate model 2 may be 
written in the same manner.  The Bayes factor is then
\begin{equation*}
    BF_{12} = \frac{f_1(\bm{Y})}{f_2(\bm{Y})},
\end{equation*}
with values greater than 1 favoring Model 1.  

Because the integral
from~\eqref{eq:ml} is generally difficult to calculate,
\cite{lewraf97} describe a Laplace-Metropolis estimator of the Bayes
factor \citep[also see][]{raf93}.  This estimator relies on the Laplace approximation to
integrals that involve a natural exponent.  Such integrals can be
written as 
\begin{equation*}
    \displaystyle \int \exp(h(\bm{u})) \partial \bm{u} \approx
    (2\pi)^{Q/2} |\bm{H}^* |^{1/2} \exp(h(\bm{u}^*)),
\end{equation*}
where $h()$ is a function of the vector $\bm{u}$, $Q$ is
the length of $\bm{u}$, $\bm{u}^*$ is the value of $\bm{u}$ that
maximizes $h$,  and $\bm{H}^*$ is the inverse of the information
matrix evaluated at $\bm{u}^*$.  As applied to marginal
likelihoods, we take $h(\bm{\vartheta}) = \log(f(\bm{\vartheta}) f(\bm{Y} |
\bm{\vartheta}))$ so that our solution to Equation~\eqref{eq:ml} is
\begin{equation}
  \label{eq:lmll}
    f_1(\bm{Y}) \approx (2\pi)^{Q/2} |\bm{H}^* |^{1/2} f(\bm{\vartheta}^*)
    f(\bm{Y} | \bm{\vartheta}^*),
\end{equation}
where $\bm{\vartheta}^*$ is a vector of posterior means (or other posterior
estimate of central tendency), obtained from the MCMC output.
%% This is especially simple because $f(\bm{Y} | \bm{\vartheta}^*)$ is normal
%% for models involving normal distributions; models involving
%% categorical data would require us to numerically integrate the latent variables.
In package \pkg{blavaan}, we follow \cite{lewraf97} and compute
the logarithm of this approximation for numerical stability. 

The Bayes factor is sensitive to choice of prior distribution \citep[e.g.,][]{liuait08,van10}, so researchers using the Bayes factor are advised to carefully consider their priors. \cite{kasraf95} provide popular rules of thumb for interpreting the log-Bayes factor, though the extent to which these rules are meaningful in any specific application is unclear. In general, as the log-Bayes factor increases from 0, we gain increasing support for the first model. The Bayes factor can also be motivated as the extent to which, after observing data, we should revise the prior odds of model 1 being correct versus model 2 being correct; see, e.g., \cite{kasraf95} and \cite{roumor16}.

\end{document}